\documentclass[aps,prb,groupedaddress,superscriptaddress,twocolumn]{revtex4-2}
\usepackage{amsmath,graphicx}
\usepackage{amssymb}
\usepackage{color}
\usepackage{mhchem} 

\begin{document}


\title{Absence of edge states at armchair edges in inhomogeneously strained graphene under a pseudomagnetic field}

\author{Jing-Yun Fang}
\email[]{These authors contributed equally to this work}
\affiliation{International Center for Quantum Materials, School of Physics, Peking University, Beijing 100871, China}
\affiliation{Hefei National Laboratory, Hefei 230088, China}

\author{Yu-Chen Zhuang}
\email[]{These authors contributed equally to this work}
\affiliation{International Center for Quantum Materials, School of Physics, Peking University, Beijing 100871, China}
\affiliation{Hefei National Laboratory, Hefei 230088, China}

\author{Qing-Feng Sun}
\email[]{sunqf@pku.edu.cn}
\affiliation{International Center for Quantum Materials, School of Physics, Peking University, Beijing 100871, China}
\affiliation{Hefei National Laboratory, Hefei 230088, China}
\affiliation{Beijing Academy of Quantum Information Sciences, West Bld.\#3, No.10 Xibeiwang East Rd., Haidian District, Beijing 100193, China}

\date{\today}

\begin{abstract}
Nonuniform strain in graphene can induce a pseudo-magnetic field (PMF) preserving time-reversal symmetry, generating pseudo-Landau levels under zero real magnetic field (MF). The different natures between PMF and real MF lead to the counterpropagating valley-polarized edge states under the PMF and unidirectionally chiral edge states under the real MF. In this work, we find, due to the valley mixing on the armchair edges, the quantum valley Hall edge states only exist at the zigzag edges but not at armchair edges in a uniaxial strained graphene, very different from the case that chiral quantum Hall edge states exist at all edges in pristine graphene under a real MF. We theoretically demonstrate it through the wave function distributions, multi-terminal transport measurements and the electron local occupations, respectively. The interface state in a $p$-$n$ junction under PMF is further proposed to transport electrons between the conductive zigzag boundaries, which could be used as a valley-polarized single pole double throw switch.
\end{abstract}

\maketitle

\section{\label{secS1}Introduction}
Graphene, a fascinating carbon monolayer membrane with honeycomb lattice, has stimulated extensive research interests because of its novel fundamental features and possible applications~\cite{Rev.Mod.Phys.81.109,Rev.Mod.Phys.80.1337,Nat.Mater.6.183,Science.324.1530}. It has a peculiar band structure harboring a linear dispersion relation, where the carriers behave as massless Dirac fermions~\cite{Nature.438.197}. This leads to considerable outstanding properties such as Klein paradox~\cite{Nat.Phys.2.620}, half-integer quantum Hall effect~\cite{Phys.Rev.Lett.95.146801,Science.315.1379}, weak anti-localization~\cite{Rev.Mod.Phys.82.2673}, atomic collapse states~\cite{addref1,addref2}, to name a few. One of the signature properties among these is its exceptional mechanical flexibility, which can sustain nondestructive reversible deformations up to 25\%~\cite{Science.321.385}. This allowed people to tailor the electronic structure and transport properties by mechanical deformation~\cite{Phys.Rev.Lett.103.046801,Science.336.1557,Phys.Rev.B.103.155433}, making graphene attractive for tempting prospects in practical applications at the nanoscale. A new discipline, straintronics, controlling the spectrum of materials by strain engineering, has emerged in this context.

In essence, strain can induce the lattice distortions together with local changes of Dirac fermions. This is mathematically equivalent to modulate the hopping amplitudes between neighboring lattice sites, giving rise to a spatially varying Fermi velocity~\cite{Phys.Rev.Lett.108.227205} and a synthetic gauge field. Such a synthetic gauge field further reveals a strong pseudo-magnetic field (PMF) as well as quantized pseudo-Landau levels~\cite{Phys.Rep.496.109}, very similar to a real magnetic field (MF). By tuning the deformed structure of graphene, the PMFs can be changed from zero to hundreds of Tesla \cite{Science.329.544,Phys.Rev.B.107.195417}. Therefore, graphene-based strain engineering is viewed as a potential route to realize previously unattainable novel quantum states beyond real MFs. And combined with a high carrier mobility~\cite{Nature.457.706,Science.306.666,Nat.Nanotechnol.5.722}, graphene may replace traditional silicon-based electronic technologies, as well as construct multi-functional devices by strain in the future~\cite{Nature.479.338,Nat.Nanotechnol.3.654,Phys.Rev.B.106.035416}. Motivated by the enormous application prospects, a myriad of efforts have been made to realize specially designed strained patterns and create PMFs in graphene~\cite{Nat.Phys.6.30,Phys.Rev.B.81.035408,Phys.Rev.Lett.115.245501,Nature.483.306,Nat.Commu.11.371}. For example, theoretical realizations of PMFs have been proposed through a uniaxial stretch~\cite{Phys.Rev.Lett.115.245501}, triaxial strain~\cite{Nat.Phys.6.30}, and by naturally bending graphene nanoribbon (GNR)~\cite{Phys.Rev.B.81.035408}, etc. Experimentally, such giant PMFs are accessible by observing well-spaced pseudo-Landau levels. Up to now, the experimental verifications of PMFs have been demonstrated in nanoscale graphene nanobubbles suspended on the substrate~\cite{Science.329.544}, deformed artificial molecular graphene~\cite{Nature.483.306}, twisted bilayer graphene~\cite{Nat.Commu.11.371}, and buckled graphene superlattices~\cite{Nature.584.215}, etc. 

Unlike the real MFs breaking the time-reversal symmetry, PMFs preserve the time-reversal symmetry with opposite signals at two inequivalent Dirac points (labeled as $K$ and $K'$ valleys)~\cite{Nat.Phys.6.30}. The distinct nature between PMFs and real MFs can bring interesting phenomena in GNRs. For example, a special valley-polarized Landau quantization~\cite{Phys.Rev.Lett.124.106802}, valley-polarized confined states~\cite{Phys.Rev.Lett.129.076802}, and valley-polarized currents~\cite{Phys.Rev.B.101.085423} are proposed under PMF induced by strained GNRs, which are unattainable easily in pristine GNRs in principle~\cite{Nat.Phys.3.172}. And the flat bands can exist in uniaxially periodic strained GNRs~\cite{Phys.Rev.B.107.235143}. Moreover, the PMF could also lead a sublattice polarization of the zeroth pseudo-Landau level~\cite{Phys.Rev.B.93.035456, Nature.584.215}. As the real MF can induce the quantum Hall effect in graphene~\cite{Nature.438.201}, the induced PMF, mimicking a real MF, is suggested to generate a novel topological phase: the pseudo-quantum Hall effect (known as quantum valley Hall effect)~\cite{Nat.Phys.6.30, NanoLett.10.3551,Phys.Rev.B.106.115437}. The opposite directions of PMF in $K$ and $K'$ valleys result in a pair of counterpropagating valley-polarized edge states at the boundary~\cite{Phys.Rev.B.106.165417,Phys.Rev.B.105.195420}, quite different from the chiral edge states in a real MF propagating unidirectionally at each boundary~\cite{Phys.Rev.B.73.205408}.

There are two well-known types of edges for GNRs, namely, zigzag and armchair edges [see Fig.~\ref{fig:1}(a)]. The edge configuration of GNRs affects the electronic properties by strongly influencing the $\pi$-electron states near the Fermi level~\cite{Phys.Rev.B.59.8271}. The armchair GNRs could be metallic or insulating depending on various nanoribbon widths along the armchair direction, while the zigzag GNRs are always metallic, carrying an additional flat band at charge neutral point~\cite{Phys.Rev.B.73.235411,Phys.Rev.B.54.17954}. Particularly, valleys are decoupled for zigzag GNRs but are mixed for armchair GNRs~\cite{Phys.Rev.B.73.235411}, which could also influence the energy bands of strained graphene \cite{Phys.Rev.B.87.155422,NanoRes.3.189}. Specifically, under an inhomogeneous uniaxial strain, the PMF can be created only when strain is applied on the zigzag GNRs~\cite{Chin.Phys.B.30.030504}. However, for edge-state transport, it is shown that the boundary effects can be ignored under strong real MFs~\cite{Phys.Rev.Lett.96.176803,Phys.Rev.B.80.235411}. The chiral edge states can emerge both along the armchair and zigzag edges under a strong perpendicular real MF. Naturally, an interesting question arises: whether the edge states could also survive at both edges under the strain-induced PMF. To the best of our knowledge, many previous studies have observed the counterpropagating edge states along the zigzag edges caused by PMF in strained graphene~\cite{NanoLett.10.3551,Phys.Rev.B.105.195420}, but few studies focused on the edge-state transport along armchair edges in strained graphene.

As building blocks, we show that no conducting edge states exist at the armchair edges in strained graphene in this paper [see Fig.~\ref{fig:2}(c)]. Since a uniaxial strain applied on the zigzag GNRs breaks the translation invariance along the armchair edge, the energy band analysis adapted in periodic structure is no longer applicable at the armchair edge. Thus, we parallelly conduct the analysis from three ways. First, we give the wave function distributions of the edge states in a rectangular-shaped uniaxial strained graphene nanoflake, which can help us study the states at armchair edges. We find the edge states are distributed only at the zigzag edges without any states at the armchair edges. This is in sharp contrast to the consensus that circling edge states exist in pristine graphene nanoflakes subjected to a real MF (see Fig.~\ref{fig:3}). Second, we give the results of edge-state transport measurements in four-terminal and two-terminal devices. We find the transmission coefficients along the armchair edges are all zero in strained graphene, strongly indicating that there is no conducting edge state at the armchair edges. Third, we study the local occupation of electrons in two-terminal strained GNRs with a small bias. Driven by the small bias, we find that electrons only propagate along the zigzag edges and eventually return to the same terminal under PMF, but never propagate from one terminal to another along the armchair edges. This also vividly demonstrates the absence of conductive channels at the armchair edges in strained graphene. Moreover, we construct a $p$-$n$ junction based on strained graphene and find that snake states can flow in the interface of the $p$-$n$ junction. Therefore, electrons can propagate from one zigzag edge to another through the interface states, which is unattainable in the individually strained graphene due to the insulation of armchair edges. Considering this special characteristic, we further propose a single pole double throw switch tuned by a local gate voltage. This multi-way switch is valley-polarized, thus can also act as a valley filter \cite{Nat.Phys.3.172}.

The rest of the paper is organized as follows.
In Sec.~\uppercase\expandafter{\romannumeral2},
we describe the model of an inhomogeneous strained graphene in which a uniform PMF is obtained by applying a uniaxial strain. The energy bands of GNRs under the PMF and the real MF are also shown in this section. In Sec.~\uppercase\expandafter{\romannumeral3}, using a numerical exact diagonalization of the tight-binding model, we give the wave function distributions of the edge states in rectangular-shaped graphene nanoflakes, which clearly show that no conducting edge states exist at the armchair edges. In Sec.~\uppercase\expandafter{\romannumeral4}, combining Landauer-B\"{u}ttiker formalism and nonequilibrium Green's function method, the insulation of armchair edges is verified by zero transmission coefficients in the multi-terminal devices. In Sec.~\uppercase\expandafter{\romannumeral5}, we take up the study of local occupation number, which directly shows that there is absence of edge states at armchair edges. Considering that electrons can only propagate along the zigzag edge in individually strained graphene, a $p$-$n$ junction under the strain-induced PMF to transport electrons from one zigzag edge to another through interface snake states is designed in Sec.~\uppercase\expandafter{\romannumeral6}. A single pole double throw switch as well as a valley filter tuned by a local gate voltage are also proposed. And a brief conclusion is presented in Sec.~\uppercase\expandafter{\romannumeral7}. Besides, we give an Appendix A to show the wave function centers for states in the rectangular-shaped graphene nanoflakes in detail.

\section{\label{secS2}Strained graphene Hamiltonian and energy bands}
\begin{figure}
	\includegraphics[scale=0.32]{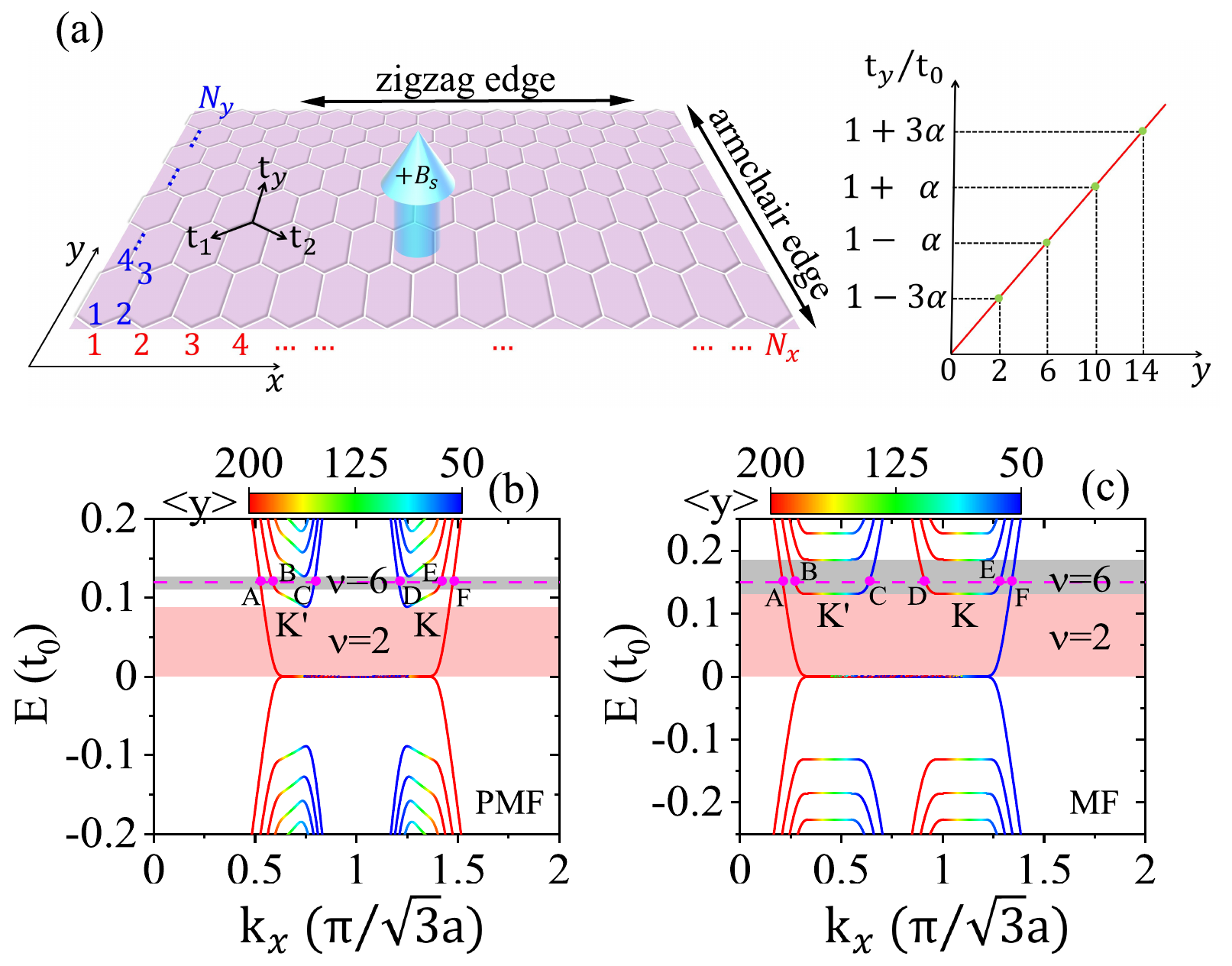}
	\centering
	\caption{(a) Schematic illustration of realizing a uniform PMF $B_s$ in strained graphene by tuning the hopping strength $t_y$ along the $y$-axis denoting by black arrows. The number $1,2,3,4...$ marked by red and blue show the lattice indices $(x,y)$ of carbon atoms along the $x$ and $y$ direction. They can be transformed into the actual carbon atom positions $(\tilde{x},\tilde{y})$ in the unit of nm. 
	Ignoring the small effect of strain on lattice positions, 
	$(\tilde{x},\tilde{y})=(\sqrt{3}(x-1)a, \frac{3(y-1)a}{4})$, $(\sqrt{3}(x-1)a+\frac{\sqrt{3}}{2}a, \frac{3(y-2)a}{4}+\frac{a}{2})$, $(\sqrt{3}(x-1)a+\frac{\sqrt{3}}{2}a, \frac{3(y-3)a}{4}+\frac{3a}{2})$, and $(\sqrt{3}(x-1)a, \frac{3(y-4)a}{4}+2a)$ for $y\mod4 =1$, $2$, $3$, and $4$, respectively. Besides, the corresponding strain strength $\tilde{\alpha}$ in the unit of nm$^{-1}$ is $\tilde{\alpha} \approx \frac{4\alpha}{3a}$. As an example, the plot on the right shows schematically the hopping strength $t_y$ vs even lattice index $y$ with strain parameter $\alpha$ for $N_y=16$. (b) and (c) are energy bands for strained zigzag GNR with $\alpha=0.005$ and pristine zigzag GNR with $\phi=-0.005$, respectively. The color scale represents the expectation value of the lattice index operator $\langle y\rangle $ for each eigenstate. The horizontal pink and gray stripes mark two different regions, labelled as filling factor $\nu=2$ and $\nu=6$, respectively.}
	\label {fig:1}
\end{figure}

First of all, we introduce the effective Hamiltonian of the inhomogeneous strained graphene. Its discrete form in the tight-binding representation can be written as~\cite{Chin.Phys.B.30.030504}
\begin{eqnarray} \label{eq:1}
H=\sum_{\bf i}\varepsilon_{\bf i}c_{\bf i}^{\dag}c_{\bf i}+\sum_{<\bf ij>}t_{\bf ij}e^{i\phi_{\bf ij}}c_{\bf i}^{\dag}c_{\bf j},
\end{eqnarray}
where $\varepsilon_{\bf i}$ is the on-site energy, and $c_{\bf i}^{\dag} (c_{\bf i})$ is the creation (annihilation) operator. When considering a real MF $\boldsymbol{B}$, a phase $\phi_{\bf ij}=\int_{\bf i}^{\bf j}{\boldsymbol{A}\cdot{d\boldsymbol{l}}/\phi_{0}}$ is added in the hopping element with the vector potential $\boldsymbol{A}=(-By,0,0)$ and $\phi_{0}=\hbar/e$ the flux quantum \cite{Phys.Rev.B.104.115411}. As for the PMF $\boldsymbol{B_s}$, we consider an inhomogeneous uniaxial strain along the armchair ($y$) direction [see Fig.~\ref{fig:1}(a)]. The bond length between the adjacent lattice sites along the $y$ direction decreases from the lower edge to the upper edge, which effectively renormalize the hopping parameter $t_{\bf ij}$ by an exponential relation \cite{Phys.Rev.B.80.045401}. In our case, the hopping coefficient $t_{\bf ij}$ can be simply classified into two categories [denoted by black arrows in Fig.~\ref{fig:1}(a)]: (i) the hopping parameter along $y$ direction $t_{y}$, (ii) the hopping parameters along the other two main crystallographic directions $t_{1}$, $t_{2}$. In $y$ direction, the hopping coefficient is approximately set to a linear function dependent on the position $y$ with $t_y=t_0[1+\frac{\alpha}{2}(y-\frac{N_{y}}{2})]$~\cite{Phys.Rev.B.105.195420, Chin.Phys.B.30.030504}. Here $y$ denotes the even lattice index along y direction and $\alpha$ is a tunable parameter reflecting the strain strength [see Fig.~\ref{fig:1}(a)]. $N_y$ is the width of the zigzag GNRs, and $t_0=2.75$ eV is the well-known hopping coefficient as the energy unit. A typical value distribution of $t_y$ is schematically shown by the right plot in Fig.~\ref{fig:1}(a) with $N_y=16$. In fact, the hopping parameter $t_{1}$, $t_{2}$ in the other two main crystallographic directions should also change under the uniaxial deformation along y direction. However, their changes are much smaller than $t_{y}$, so we can simply set them as a constant with $t_1=t_2=t_0$~\cite{Phys.Rev.B.103.155433,Phys.Rev.B.105.195420,Phys.Rev.B.80.045401,Chin.Phys.B.30.030504}. This uniaxial strain is theoretically investigated by stretching graphene along a certain direction~\cite{Phys.Rev.Lett.115.245501}. However, we emphasize that the absence of armchair edge states is in principle general for the PMF realized by various methods~\cite{Science.329.544,Science.336.1557,Nature.483.306,Nat.Commu.11.371,Nature.584.215}, considering the PMF is opposite at $K$ and $K'$ valleys. Our strain pattern breaks the inversion symmetry of the lattice, which is a necessary ingredient to generate a PMF~\cite{Nat.Phys.6.30}. To evaluate the induced PMF, we can express the effective Hamiltonian of graphene at low energy in the momentum space~\cite{Chin.Phys.B.30.030504}
\begin{eqnarray} \label{eq:2}
	H_{\boldsymbol{k}}=& \nonumber \\ 
	&\begin{pmatrix}0 & t_y+t_1e^{i\boldsymbol{k}\cdot \boldsymbol{a}_{1}}+t_2e^{i\boldsymbol{k}\cdot \boldsymbol{a}_{2}} \\ t_y+t_1e^{-i\boldsymbol{k}\cdot \boldsymbol{a}_{1}}+t_2e^{-i\boldsymbol{k}\cdot \boldsymbol{a}_{2}} & 0\end{pmatrix},  \nonumber \\
\end{eqnarray}\\
where $\boldsymbol{a}_1=(\frac{\sqrt{3}a}{2},\frac{3a}{2})$, 
$\boldsymbol{a}_2=(-\frac{\sqrt{3}a}{2},\frac{3a}{2})$ are primitive vectors and $a=0.142$ nm is the distance between two nearest-neighbor carbon atoms. For pristine graphene, $t_y=t_1=t_2=t_0$. 
Satisfying the equation $t_0+t_0e^{i\boldsymbol{K}_{\rm D}\cdot \boldsymbol{a}_{1}}+t_0e^{i\boldsymbol{K}_{\rm D}\cdot \boldsymbol{a}_{2}}=0$, 
we can pick up one Dirac point $\boldsymbol{K}_{\rm D}$ in $K$ valley with $\boldsymbol{K}_{\rm D}=(K_{\rm D}^{x},K_{\rm D}^{y})=(\frac{4\pi}{3\sqrt{3}a},0)$. After introducing the above uniaxial strain in the graphene, the new Dirac point is shifted from $\boldsymbol{K}_{\rm D}$ to $\boldsymbol{K}_{\rm S}=\boldsymbol{K}_{\rm D}+\boldsymbol{\delta}$, which should still satisfy the equation $t_y+t_1e^{i\boldsymbol{K}_{\rm S}\cdot \boldsymbol{a}_{1}}+t_2e^{i\boldsymbol{K}_{\rm S}\cdot \boldsymbol{a}_{2}}=0$. Considering a small shift for $\boldsymbol{\delta}=(\delta_x,0,0)$, we have 
\begin{eqnarray} \label{eq:3}
	t_0[1+\frac{\alpha}{2}(y-\frac{N_{y}}{2})]+2t_0{\rm cos}[\frac{\sqrt{3}a}{2}(K_{\rm D}^{x}+\delta_x)]=0.
\end{eqnarray}
 By solving the Eq.~(\ref{eq:3}) through Taylor expression at Dirac point $\boldsymbol{K}_{\rm D}$,  $\delta_x$ can be obtained,
then the strain-induced pseudo-gauge field at $K$ valley 
is derived as $\boldsymbol{A_s}=(A_x,0,0)$ with $A_x=\frac{\hbar}{e}\delta_x=\frac{\hbar\alpha }{3ea}(y-\frac{N_y}{2})$. Similarly, the pseudo-gauge field at $K'$ valley is $-\boldsymbol{A_s}$ due to time-reversal symmetry. And a uniform perpendicular PMF can be obtained by $\boldsymbol{B_s}=\nabla \times \boldsymbol{A_s}=(0,0,B_s)$ with $B_s=\mp \frac{\hbar\alpha}{3ea}$. In the following calculations, we will take $\alpha=0.005$ ($B_s \approx \mp  73$ T) and $N_y=280$, thus the maximum deformations are $0.655t_0$ and $1.345t_0$ in the lower and upper edge positions, which can be achieved nondestructively in experiments~\cite{Science.321.385,Phys.Rev.Lett.102.235502}. Under this set of parameters, the pseudo-Landau levels can be well formed. For the real MF, we employ the Peierls phase: $2\phi=(3\sqrt{3}/2)a^2B/\phi_{0}$, with $(3\sqrt{3}/2)a^2B$ the magnetic flux threading a single hexagon and $\phi_{0}=\hbar/e$. In the following calculations, we will take $\phi=-0.005$ and the Landau levels can be well formed under this parameter. We address the results are independent of the specific value of strain parameter $\alpha$ and the real MF strength $\phi$, as long as the (pseudo-)Landau levels as well as the corresponding edge states can well survive. 

When strain is applied along the $y$ direction, the translation invariance along the $y$ direction is broken but preserved along the $x$ direction. Then we can still calculate the spectrum of the strained zigzag GNR. Figure~\ref{fig:1}(b) shows the band structure of a strained zigzag GNR with strain parameter $\alpha=0.005$, in which we also give color maps to denote the averaged lattice location $\langle y\rangle $ of the eigenstates. We can find that PMF acts on the electrons leading to the dispersive pseudo-Landau levels. Due to modulated inhomogeneous Fermi velocities~\cite{Phys.Rev.B.101.085423} as well as time-reversal symmetry, the pseudo-Landau levels are tilted with opposite slopes in the two valleys near the Dirac points. The zeroth pseudo-Landau level is flat since it is independent of the Fermi velocity. In Fig.~\ref{fig:1}(b), we also denote the region of filling factor $\nu=2$ (pink region) and $\nu=6$ (gray region), respectively. Note that a gap between them is just the range for the tilted first pseudo-Landau level. In this gap, the dispersive bulk states and edge states will coexist.  From the color distributions on the band structure, we can see that electrons propagate in opposite directions coming form different valleys at each zigzag edge. That is to say, the edge states are valley-polarized and valley-momentum locked, which can lead to the quantum valley Hall effect in strained graphene~\cite{NanoLett.10.3551,Nat.Phys.6.30,Phys.Rev.B.106.165417}. For comparison, we also give the band structure of a pristine zigzag GNR subjected to a real MF, see Fig.~\ref{fig:1}(c). It is clear that the well-defined $n$-th flat Landau level locates at $E_n=\text{sign}(n)\sqrt{2e\hbar v_xv_yB\left\lvert n \right\rvert}$ ($n=0, \pm 1, \pm 2, ...$) in the low-energy regime with $v_x=v_y$ the Fermi velocity for $x$ and $y$ directions, respectively, which is consistent with Refs.~\cite{Chin.Phys.B.30.030504,Phys.Rev.B.106.165417}. Different from the case of PMF, all Landau levels are totally flat under the real MF and there is no gap between the regions of $\nu=2$ and $\nu=6$. From the color distributions on the band structure, we can see that electrons with the same group velocity direction are located at the same edge. In other words, the edge states are chiral under the real MF. The chiral edge states can lead to the quantum Hall effect in graphene, which have been shown in many previous works~\cite{Nature.438.197,Nature.438.201,Phys.Rev.Lett.95.146801,Phys.Rev.B.104.115411}.

\begin{figure*}
	\includegraphics[scale=0.4]{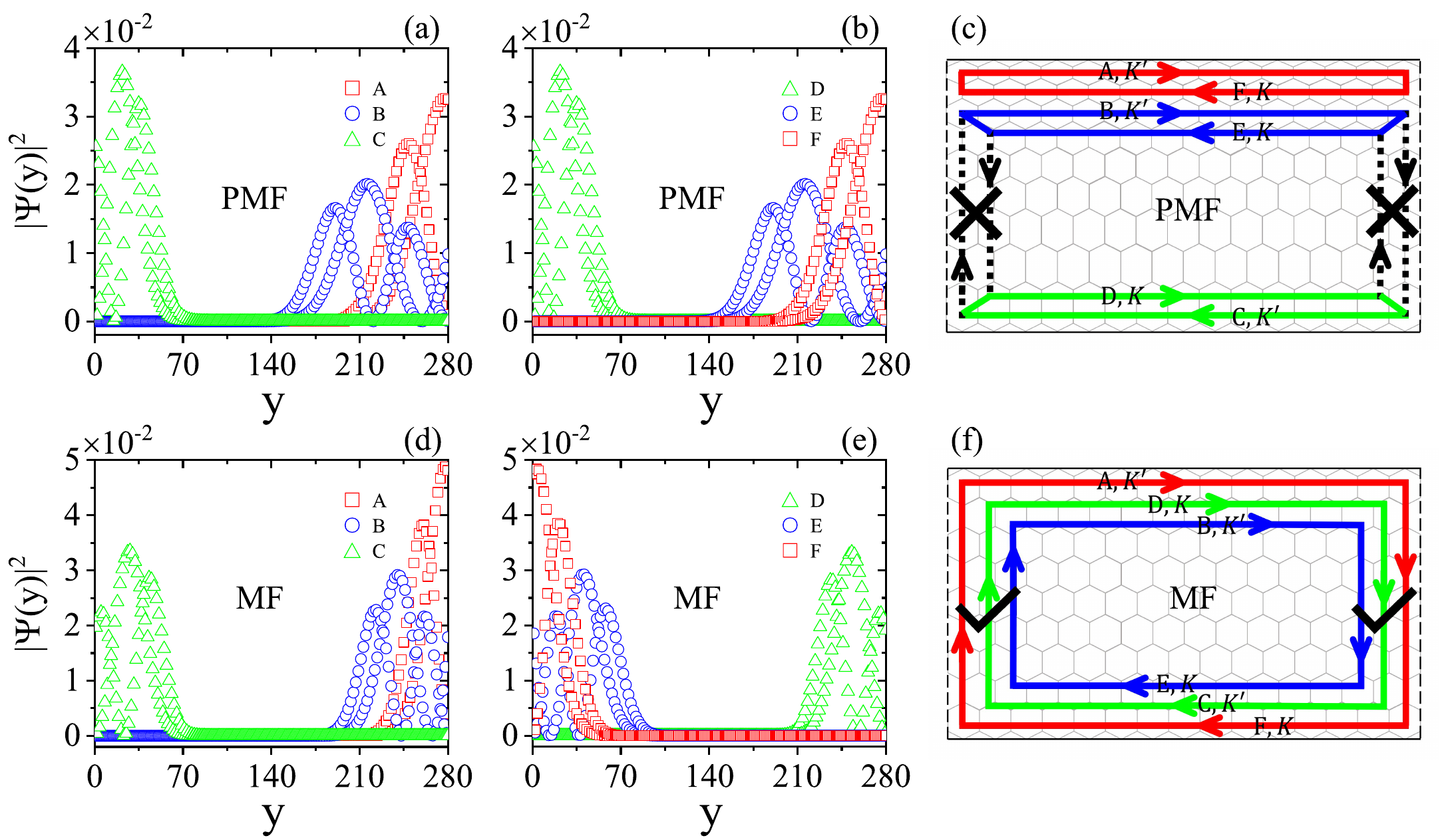}
	\centering
	\caption{(a),(b) [(d),(e)] The wave function distributions $|\Psi(y)|^2$ vs vertical lattice index $y$ and (c) [(f)] schematic diagram for the edge modes $A$-$F$ marked by magenta dots in Fig.~\ref{fig:1}(b) [Fig.~\ref{fig:1}(c)] for PMF (real MF), with the Fermi energy $E_F=0.12t_0$ ($0.15t_0$). In (a)-(c), the edge states in pairs are valley-polarized, and electrons propagate in the opposite direction for different valleys at each edge. In (d)-(f), the edge states are chiral, electrons propagate in the same direction at the same edge.}
	\label {fig:2}
\end{figure*}

In order to visualize the distributions of edge states belonging to each band, we give the wave functions of the edge states corresponding to the filling factor $\nu=6$ directly, approximately with the range $E \in [0.11t_0,0.13t_0]$ for the PMF and $E \in [0.13t_0, 0.19t_0]$ for the real MF [see the gray region in Figs.~\ref{fig:1}(b) and \ref{fig:1}(c)]. Without loss of generality, we select the Fermi energy $E_F=0.12t_0$ in Fig.~\ref{fig:1}(b) and $E_F=0.15t_0$ in Fig.~\ref{fig:1}(c), see the magenta dotted lines. The 6 edge modes, i.e., the intersections of the selected Fermi energy $E_F$ and the band, are labeled as $A$-$F$ from the left to right. The corresponding distributions of wave functions for these edge states are shown in Fig.~\ref{fig:2}. Figures~\ref{fig:2}(a) and \ref{fig:2}(b) first show the distributions of wave functions for the edge states belonging to $K'$ and $K$ valley in strained graphene, respectively. Unlike the real MF, PMF preserves the time-reversal symmetry, thus the wave functions in $K'$ and $K$ valleys are similar, i.e., modes $A$ and $F$, $B$ and $E$, $C$ and $D$ have the same distributions. Besides, the edge states at the upper and lower zigzag edges are asymmetric under the PMF due to broken inversion symmetry. Two additional edge states $A$ and $F$ associated with the zeroth pseudo-Landau level ($\nu=2$) are located only near the upper edge. Thus, the upper edge always has two more counterpropagating edge states than the lower one under PMF. Combining the energy bands and the distributions of wave functions, we further sketch the flow of edge states roughly, see Fig.~\ref{fig:2}(c). Arising from the time-reversal symmetry, there are edge states propagating in opposite directions along both the upper and lower edges. Besides, such counterpropagating edge states with opposite chirality appear in pairs belonging to $K$ and $K'$ valley at each edge, respectively. That is to say, the edge states in PMF are valley-polarized and valley-momentum locked, so pure valley currents are possible in this case~\cite{Phys.Rev.Lett.124.106802,Phys.Rev.B.105.195420}. In Figs.~\ref{fig:2}(d) and \ref{fig:2}(e), we also give the distributions of wave functions for the edge states belonging to $K'$ and $K$ valley in pristine graphene with a real MF. We can see that the wave functions associated with valley $K'$ present a symmetric relationship with those in valley $K$, i.e., modes $A$ and $F$, $B$ and $E$, $D$ and $C$ are symmetrically distributed near the upper and lower zigzag edges. The results are also consistent with the energy band in Fig.~\ref{fig:1}(c) and can be attributed to inversion symmetry of the system. We also show the flow of the edge states for the pristine GNR with a real MF, see Fig.~\ref{fig:2}(f). The edge states are chiral, electrons propagate in the same directions at the same edge.

The above analysis illustrates the edge states at the upper and lower zigzag edges in zigzag GNRs both for PMF and real MF. Now we turn to the armchair edges. Due to the different band structures of the armchair edge and the zigzag edge, different physical phenomena are expected. To be worth mentioned, previous studies have shown that the effect of graphene boundary can be ignored in strong real MFs, and the chiral edge states well survive at the armchair edges in pristine GNRs under a real MF~\cite{Phys.Rev.Lett.96.176803,Phys.Rev.Lett.101.166806}. Therefore, for an isolated pristine graphene nanoflake under real MFs, the edge states at the upper and lower zigzag edges will be connected by those at the armchair edges, and form closed loops together in which electrons flow along the same direction (clockwise here), see Fig.~\ref{fig:2}(f). However, the translation invariance along the armchair edges is broken in the inhomogeneous uniaxial strained GNR, making it impossible to directly calculate the energy band at armchair edges. This in some sense makes edge states along the armchair edges in strained graphene little attention. To the question of whether there are edge states at the armchair edges in strained graphene, Our answer is no [see Fig.~\ref{fig:2}(c)]. In the following sections, we will illustrate this intriguing result.

\section{\label{secS3}Wave function distributions in graphene nanoflakes}

\begin{figure*}
	\includegraphics[scale=0.5]{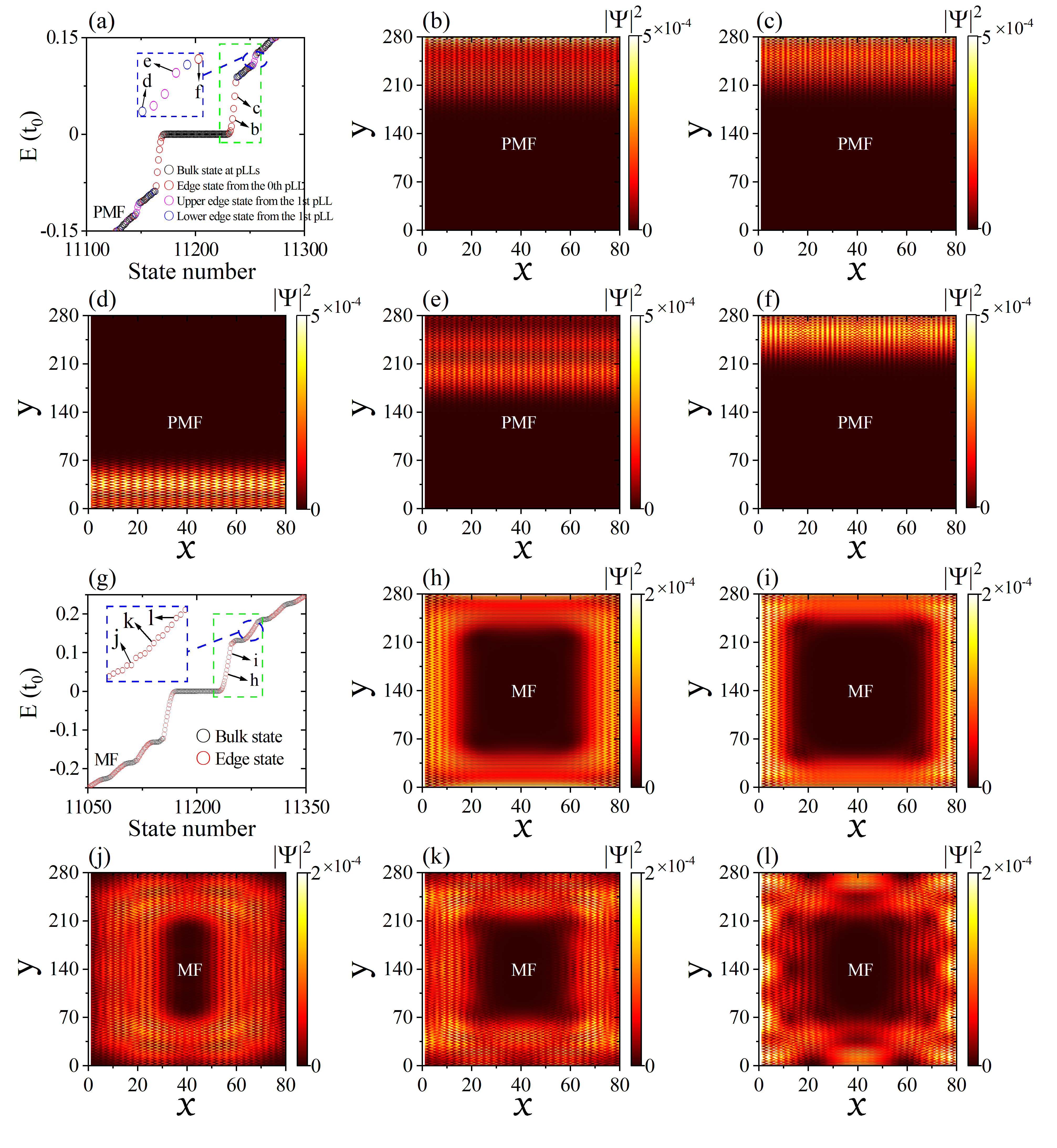}
	\centering
	\caption{(a) [(g)] Eigenvalues from solving the real-space lattice Hamiltonian of a rectangular-shaped strained (pristine) graphene nanoflake under the PMF (real MF), with open boundary conditions both in the $x$ and $y$ directions. The sample size is $N_x\times N_y=80\times280$ (about 20 nm $\times$ 30 nm). The word pLL in the legend stands for pseudo-Landau level. The inset in (a) [(g)] is the corresponding enlarged figure for the edge modes with $\nu=6$. The region surrounded by the green dashed lines in (a) [(g)] is also shown in Fig.~\ref{fig:A1}(a) [Fig.~\ref{fig:A1}(d)]. The density profiles of corresponding edge modes $b$-$f$ ($h$-$l$) marked in (a) [(g)] vs lattice indices $(x,y)$ are shown in (b)-(f) [(h)-(l)], respectively. One can see that the density profiles of the edge modes are distributed only at the upper or lower zigzag boundaries under the PMF in (b)-(f), with no closed loop connecting the armchair and zigzag edges. While the density profiles of the edge modes are all distributed over the whole boundary forming closed loops under the real MF in (h)-(l).}
	\label {fig:3}
\end{figure*}

In this section, we focus on the case of edge states on the armchair edges. Since the introduced inhomogeneous uniaxial strain breaks translation invariance at the armchair edge, the analyses of energy bands and edge states as shown in Fig.~\ref{fig:1} in a periodic structure are failed on the armchair edges. Considering this point, we change to a finite nanoflake. Using a numerical solution of the tight-binding model in Eq.~(\ref{eq:1}), we calculate the eigenvalues of the states in a rectangular-shaped graphene nanoflake with exact diagonalization, which allows us to obtain the spatial distribution of wave functions of all the states. Figure~\ref{fig:3}(a) shows the eigenvalues of the graphene nanoflake under PMF induced by strain. These eigenvalues are approximately consistent with the energy dispersion calculated in Fig.~\ref{fig:1}(b). We can clearly see the bulk states at the flat zeroth pseudo-Landau level with zero energy and the higher tilted pseudo-Landau levels with nonzero slope, denoted by black circles in Fig.~\ref{fig:3}(a). Between the neighboring pseudo-Landau levels, the edge states appear with a larger slope than the tilted pseudo-Landau levels. To be precise, we further distinguish these edge states from their density profiles and also their wave functions centers (See Appendix A for more details). Here the red circles represent the edge states originating from the zeroth pseudo-Landau level, and the magenta and blue circles represent the upper and lower edge states originating from the first pseudo-Landau level, while those with higher energy are not shown here [see legends in Fig.~\ref{fig:3}(a)]. Interestingly, we can find these edge states can coexist with the bulk states at tilted pseudo-Landau levels. For example, see the blue, red circles and black circles in the range of the first pseudo-Landau level ($E\in[0.09t_0,0.11t_0]$) in Fig.~\ref{fig:3}(a) and their enlarged figure in Appendix A, which roughly corresponds to the gap region between $\nu=2$ and $\nu=6$ [see Fig.~1(b)].

In the following, we study the wave function distributions $|\Psi\vert^{2}$ of the edge states vs lattice indices $(x,y)$ under PMF. For the filling factor $\nu=2$ where the edge states located between the zeroth pseudo-Landau level and the first pseudo-Landau level, we select two eigenstates without any specificity, see the signals $b$ and $c$ marked in Fig.~\ref{fig:3}(a). The density profiles of edge modes $b$ and $c$ are shown in Figs.~\ref{fig:3}(b) and \ref{fig:3}(c), respectively. Different from the bulk states distributed throughout the pristine graphene nanoflakes without PMF, we can find the local density of states apparently distribute only at the upper zigzag boundary (bright regions). They correspond to a pair of counterpropagating edge states associated with the zeroth pseudo-Landau level when $\nu=2$ in Fig.~\ref{fig:1}(b). Such an asymmetric edge state distribution can be understood in two perspectives: (i) strain breaks the inversion symmetry of the graphene, as well as the corresponding quantum Hall edge states. (ii) Time-reversal symmetry always guarantees that Kramers pairs appear at the same position with opposite velocities. In this case, since the lower boundary has no edge states, electrons traveling along one edge channel at the upper zigzag edge are scattered at the corners by the armchair edge, and then they can only back to the reverse edge channel at the same boundary. Thus the density profiles naturally distribute at the upper zigzag boundary for $\nu=2$.

For filling factor $\nu=6$, more edge modes emerge. We select edge modes $d$-$f$ without any specificity [enlarged figure surrounded by blue dashed lines in Fig.~\ref{fig:3}(a)]. Their local density profiles are shown in Figs.~\ref{fig:3}(d)-\ref{fig:3}(f), respectively. Like Figs.~\ref{fig:2}(a) and \ref{fig:2}(b), we can see that the edge states survive at the upper or lower zigzag edges. Actually, the shape and space broadening for vertical line cuts of the wave functions in Figs.~\ref{fig:3}(d)-\ref{fig:3}(f) are also very similar to the coupled edge modes $C$ and $D$, $B$ and $E$, $A$ and $F$ in Figs.~\ref{fig:2}(a) and \ref{fig:2}(b), respectively. Intuitively, these edge states at the upper and lower zigzag edges could be connected together through the edge states at the left and right armchair boundaries to form closed loops in the rectangular-shaped graphene nanoflake [see the black dashed lines in Fig.~\ref{fig:2}(c)], just as those in real MF shown in Fig.~\ref{fig:2}(f). However, the calculation results in Figs.~\ref{fig:3}(d)-\ref{fig:3}(f) show that even though edge modes $d$-$f$ are distributed at the upper or lower zigzag edges, there is no closed loop to connect the armchair and zigzag edges. This strongly indicates that the armchair edges (the left and right sides of the graphene nanoflake) are insulated without any edge state in graphene under the PMF. In addition, we can see that the wave function distributions present a periodic bright and dark interference fringes, which is a characteristic of standing wave. This is because the edge states at the zigzag edges could be scattered at the corners due to the finite size effect of graphene nanoflakes, thus the counterpropagating edge states will be coupled together and form resonant states at the zigzag edges. Going deeper, the formation of quantum Hall effect is attributed to the Lorentz force caused by the real MF~\cite{NatRevPhys.2.397}. Electrons in the bulk move on cyclotron orbits leading to discrete Landau levels, while electrons at the edges bounce off edges to form skipping orbits (edge states). In PMF, the opposite pseudo-Lorentz forces for electrons from $K$ and $K'$ valley indeed form closed orbits in the bulk and pseudo-Landau level quantization [see Fig.~\ref{fig:1}(b) and Fig.~\ref{fig:3}(a)], but skipping orbits only appear at zigzag boundaries and disappear at armchair boundaries, as shown in Fig.~\ref{fig:2}(c), which is quite abnormal. Obviously, this phenomenon is related to the characteristic of graphene boundaries, and will be illustrated further.

To better illustrate the absence of edge states at armchair edges under the PMF, in Fig.~\ref{fig:3} we also give the results of a pristine graphene nanoflake subjected to a real MF for comparison. Similarly, we first calculate the eigenvalues of the pristine graphene nanoflake under the real MF. Under the real MF, the Landau levels are totally flat and the wave function centers of all edge states are similar (see Appendix A for more details), thus we only distinguish the bulk (black circles) and edge states (red circles), see Fig.~\ref{fig:3}(g). In Figs.~\ref{fig:3}(h)-\ref{fig:3}(l), we show the wave function distributions of those edge states marked by $h$-$l$ in Fig.~\ref{fig:3}(g). The vertical line cuts of their density profiles are still basically similar to modes $A$-$F$ in Figs.~\ref{fig:2}(d) and \ref{fig:2}(e) but with a coupling induced by the finite size effect. For the graphene nanoflake subjected to a perpendicular real MF, the chiral edge states have been widely confirmed existing both at the zigzag and armchair edges~\cite{,Phys.Rev.B.59.8271,Phys.Rev.Lett.96.176803,NewJ.Phys.25.033001}. And electrons travelling as skipping orbits will naturally form closed propagation loops along the graphene nanoflake boundary under the real MF, just as shown in Fig.~\ref{fig:2}(f). Our results clearly show the closed loops in the rectangular-shaped graphene nanoflake under a real MF, which are in sharp contrast with the results under a PMF. For the filling factor $\nu=2$ in Figs.~\ref{fig:3}(h) and \ref{fig:3}(i), only one edge state propagates with a relatively narrower loop [see the pink region with $\nu=2$ in Fig.~\ref{fig:1}(c)]. For the higher filling factor, the results are similar but the closed loop will be wider, since the number and space broadening of edge states become larger, see Figs.~\ref{fig:3}(j)-\ref{fig:3}(l) with $\nu=6$. 

The insulation of armchair edge under PMF can be simply illustrated by boundary conditions on valleys brought by edge configurations. It has been pointed out that the valley isospin direction (the two-component spinor of valley degree in Bloch sphere) is in the $\pm z$ direction ($K$ or $K'$) for the zigzag edge but in $x$-$y$ plane for the armchair edge~\cite{Phys.Rev.B.73.235411,Phys.Rev.B.76.035411}. At zigzag edges, the valleys $K$ and $K'$ are fully separate without any coupling. Therefore, for the edge states from $K$ and $K'$ valleys, even if they have opposite chiralities, they will not interfere with each other and always exist in real MF $B$ and PMF $B_{s}$. However, the armchair edges, acting like a in-plane magnetic moment, tend to mix the edge states from different valleys. For real MF $B$, the edge states for valleys $K$ and $K'$ both have the same chirality and thus they are preserved. For PMF $B_{s}$, since the edge states from valleys $K$ and $K'$ have the opposite chiralities due to time-reversal symmetry, the mixing of $K$ and $K'$ valleys will couple the skipping orbits with opposite directions, and thus the edge states are broken, leading to the insulation of armchair edges.

\section{\label{secS4}Transmission coefficients of multi-terminal devices}

\begin{figure*}
	\includegraphics[scale=0.4]{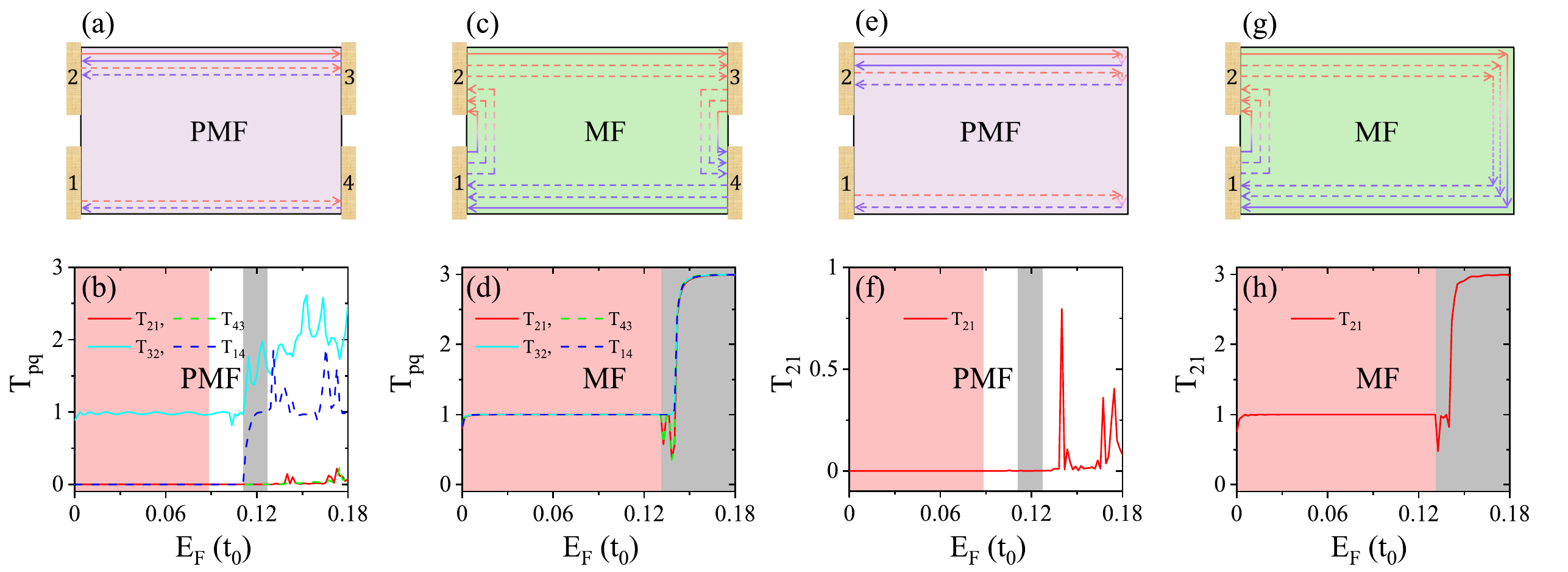}
	\centering
	\caption{(a),(c) [(e),(g)] Schematic diagrams for edge states in four-terminal (two-terminal) devices based on strained graphene nanoflakes under the PMF and pristine graphene nanoflakes under the real MF, respectively. In (a),(e), the edge states propagate in opposite directions for $K$ and $K'$ valleys at each edge, while in (c),(g), the edge states propagate unidirectionally on a given edge. (b),(d) [(f),(h)] show the corresponding transmission coefficients $T_{pq}$ ($T_{21}$) versus Fermi energy $E_F$ with terminal width $N_l=80$ for the devices in (a),(c) [(e),(g)], respectively. The pink and gray regions symbol energy regions of filling factor $\nu=2$ and $\nu=6$, respectively, which are consistent with those in Fig.~\ref{fig:1}.}
	\label {fig:4}
\end{figure*}

The insulation of armchair edges in PMF can be further revealed by transport analysis. By using the non-equilibrium Green's function method, the transmission coefficient
from lead $q$ to lead $p$ can be expressed as $T_{pq}(E)=\textmd{Tr}[{\bf \Gamma}_{p}{\bf G}^r{\bf \Gamma}_{q}{\bf G}^a]$ with the retarded Green's function ${\bf G}^r(E)=[{\bf G}^a(E)]^\dag=[E {\bf I}-{\bf  H}_{\rm cen}-\sum_{p}{\bf \Sigma}_{p}^r]^{-1}$
and the linewidth function
${\bf \Gamma}_{p}(E)=i[{\bf \Sigma}_{p}^r(E)-{\bf \Sigma}_{p}^{a}(E)]$. ${\bf H}_{\rm cen}$ here is the Hamiltonian of the central region, which is marked by purple and green for graphene under PMF and MF, see Figs.~\ref{fig:4}(a,e) and \ref{fig:4}(c,g), respectively. The electrodes are marked by yellow with the width $N_{l}=80$. In order to reduce scattering at the interfaces, we adapt graphene as the electrodes. The Hamiltonian of electrodes as well as their couplings to central region are the same as Eq.~(1). However, considering that it is difficult to introduce strain on the narrow electrodes, we choose pristine graphene with $\alpha=\phi=0$ ($t_{ij}=t_{0}$) for the electrodes if the central region is strained graphene. If the central region is pristine graphene under the real MF, we also apply real MF on electrodes and choose the same Hamiltonian for the electrodes and the central region. Then, based on methods in \cite{Phys.Rev.B.23.4997,J.Phys.F.15.851}, the retarded self-energy ${\bf \Sigma}_{p}^r(E)=[{\bf \Sigma}_{p}^{a}(E)]^\dagger$ due to the coupling to lead $p$ can be calculated numerically.

Based on the above analysis of wave function distributions in strained graphene nanoflakes, Fig.~\ref{fig:4}(a) first plots the edge states propagation of a four-terminal device under PMF. Since the counterpropagating edge states must appear in pairs at the boundaries due to time-reversal symmetry, $T_{qp}\equiv T_{pq}$. Therefore, here we only show the transmission coefficients $T_{pq}$ along the clockwise direction, as shown in Fig.~\ref{fig:4}(b). For the filling factor $\nu=2$ (pink region), there is a pair of counterpropagating edge states located only at the upper edge [see the solid lines in Fig.~\ref{fig:4}(a)]. In this case, electrons coming from terminal 2 can only reach to terminal 3, and vice versa. Thus we can expect that $T_{32}=1$, and other transmission coefficients should be zero. From Fig.~\ref{fig:4}(b), we can clearly see $T_{32}\simeq 1$ and $T_{21}=T_{43}=T_{14}=0$ as we expected, see the pink region. The slight deviation of $T_{32}$ is because of a mismatch between the lattice of unstrained electrodes and the central strained graphene. When $\nu=6$ [gray region in Fig.~\ref{fig:4}(b)], the counterpropagating edge states will appear both at the upper and the lower boundaries [see solid and dashed lines in Fig.~\ref{fig:4}(a)]. In this case, if there are conducting edge states at the armchair edges, electrons from terminal 1 (3) can arrive at terminal 4 (2) as well as terminal 2 (4). And since the upper boundary has one more pair of counterpropagating edge states than the lower boundary, the transmission coefficients should satisfy $T_{21}=T_{43}=T_{14}=1$ and $T_{32}=2$. 
However, from Fig.~\ref{fig:4}(b), we can see that the transmission coefficients along the zigzag edges, $T_{32}\simeq 2$ and $T_{14}\simeq 1$,
while the transmission coefficients along the armchair edges, i.e., $T_{21}$ and $T_{43}$ are almost zero.
This means that electrons starting from terminal 4 (2) will mainly enter to terminal 1 (3), and the rest are reflected back to terminal 4 (2) due to the lattice mismatch, but never enter to terminal 3 (1). The above transmission coefficients under the PMF in the four-terminal device are in sharp contrast with those under the real MF, where closed propagation loops form along the boundary [see Fig.~\ref{fig:4}(c)], which should satisfy $T_{21}=T_{32}=T_{43}=T_{14}=T_{\nu}$ [see Fig.~\ref{fig:4}(d)] and others are zero in the ideal case~\cite{Phys.Rev.B.104.115411}. Here $T_{\nu}=1$ and $3$ for filling factor $\nu=2$ and $6$ ignoring spin, respectively. Therefore, the zero transmission coefficients $T_{21}$ and $T_{43}$ along the strained armchair boundary well demonstrate that no closed edge state at armchair edges to connect zigzag edge states.

To eliminate the effects of electrodes, we further consider a two-terminal device as shown in Fig.~\ref{fig:4}(e) for graphene under PMF. In this case, electrons originating from terminal 1 (2) can only reach to terminal 2 (1) or return back to itself, depending on whether there are conducting edge states at the armchair edges. Under the PMF, from Fig.~\ref{fig:4}(f) we can see that the transmission coefficient along the armchair edges $T_{21}$ is always zero for the strained graphene with filling factor $\nu=2$ (pink region) and $\nu=6$ (gray region), indicating that the insulation of armchair edges is robust. The finite transmission coefficient $T_{21}$ for higher energy in Fig.~\ref{fig:4}(f) may be due to the wider broadening of edge states from higher pseudo-Landau levels or due to the bulk states. For comparison, Figs.~\ref{fig:4}(g) and \ref{fig:4}(h) also give the results of the two-terminal device under the real MF. In this case, the transmission coefficients satisfy $T_{21}=T_{12}=T_{\nu}$ in the ideal case because of the chiral edge states along the armchair boundaries, which can be well seen in Fig.~\ref{fig:4}(h). In total, the transport analysis proves that there is indeed no conducting edge state at the armchair edges under the PMF, and electrons originating from one terminal will finally return back to the same terminal through the edge states at zigzag edges, as shown in Fig.~\ref{fig:4}(e).

\section{\label{secS5}Local occupation number of two-terminal device}

\begin{figure}
	\includegraphics[scale=0.31]{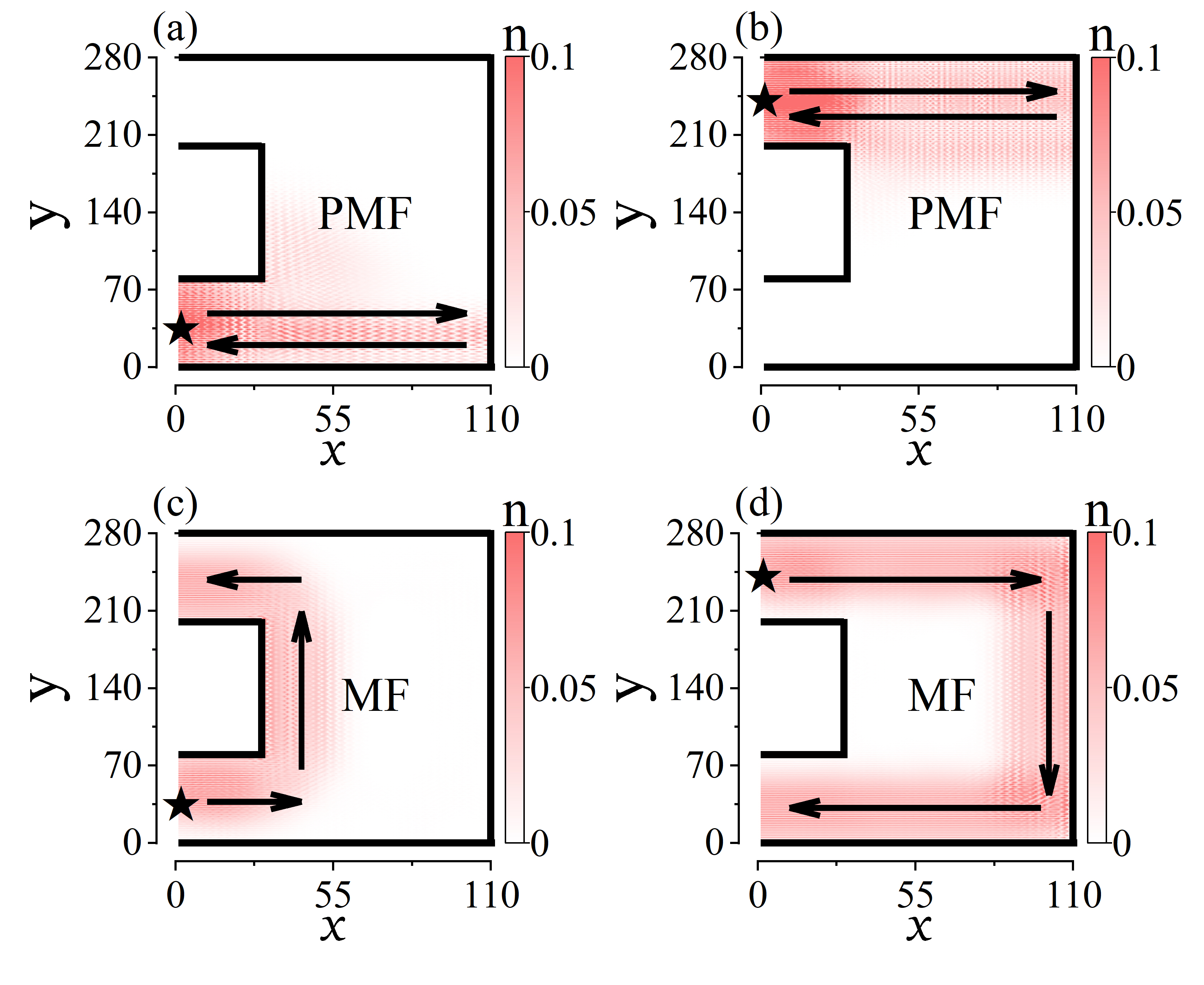}
	\centering
	\caption{The local occupation number $n$ vs lattice indices $(x,y)$ in a two-terminal device, for strained graphene nanoflakes under the PMF with $E=E_F=0.12t_0$ in (a),(b) and pristine graphene nanoflakes under the real MF with $E=E_F=0.15t_0$ in (c),(d). The black star indicates that a higher voltage is added at this terminal. And the black solid lines mark the boundary of the device while the black solid arrows indicate the propagation path of electrons. Under the PMF, electrons originating from one terminal will finally return back to the same one along the zigzag boundary. Under the real MF, electrons originating from one terminal will finally enter to another through the armchair boundary.}
	\label {fig:5}
\end{figure}

In the previous sections, we have shown the insulating armchair edges in strained graphene by the wave functions distributed only at the zigzag edges and the zero transmission coefficients along the armchair edges. To show our results more visually, we further study the flow of electrons driven by a bias in this section, which can be represented by the local occupation number. With the help of the non-equilibrium Green's function method, the total local occupation number at site $\bf i$ is given by $N_{\bf i}=\langle c_{\bf i}^{\dagger}c_{\bf i}\rangle=-i\int {\bf G}_{\bf ii}^{<}(E)dE=\int n_{\bf i}(E)dE$, where $n_{\bf i}(E)=-i{\bf G}_{\bf ii}^{<}(E)$ is the local occupation number per unit energy at site $\bf i$ with energy $E$. According to the Keldysh equation $\mathbf{G}^{<}(E)=\sum_{p}\mathbf{G}^r(E)[i\mathbf{\Gamma}_p(E)f_p(E)]\mathbf{G}^a(E)$ with $f_p(E)$ the Fermi distribution function in terminal $p$, we can rewrite $n_{\bf i}(E)$ as $n_{\bf i}(E)=\sum_{p}n_{\bf i}^{p}(E)$. Here $n_{\bf i}^{p}(E)=[\mathbf{G}^r(E)\mathbf{\Gamma}_p(E)f_p(E)\mathbf{G}^a(E)]_{\bf ii}$ is the local occupation number at site $\bf i$ for incident electrons from terminal $p$ with energy $E$. At zero temperature limit, $f_p(E)=\theta(\mu_p-E)$ where $\mu_p$ is the chemical potential in terminal $p$. Thus, the local occupation number for incident electron from the lead $p$ with energy $E \leq  \mu_{p}$ can be expressed as~\cite{Phys.Rev.B.102.245412}
\begin{eqnarray}\label{eq:4}
	n_{\bf i}^{p}(E)=[{\bf G}^r(E){\bf \Gamma}_p(E){\bf G}^a(E)]_{\bf ii}.
\end{eqnarray}

In fact, here the physical meaning of the local occupation number $n_{\bf i}^{p}$ is the same as the lead-resolved spectral function (or the lead-resolved local density of states) since the Fermi distribution function $f_p(E)=1$. Figures~\ref{fig:5}(a) and \ref{fig:5}(b) show the local occupation number of the above two-terminal device [see Fig.~\ref{fig:4}(e)] on the strained graphene through color maps, where white color represents the absence of electron occupation and red color represents the presence of electron occupation. The energy is set as $E=E_F=0.12t_0$ corresponding to $\nu=6$, thus there should be counterpropagating edge states both at the upper and lower zigzag boundaries. When a small bias is applied between terminals 1 and 2, electrons will be driven to flow from terminals with the higher voltage (higher chemical potential) to other places with lower voltage (lower chemical potential) along the edge states. In Fig.~\ref{fig:5}(a), we apply a voltage at terminal 1 (marked by the black star) and ground terminal 2. We show $n_{\bf 
i}^1$ and find that electrons are mainly distributed at the lower zigzag boundary. Similarly, in Fig.~\ref{fig:5}(b), when the higher voltage is applied at terminal 2 (marked by the black star) and terminal 1 is grounded, we show $n_{\bf 
i}^2$ and find electrons are distributed mainly at the upper zigzag boundary. This means that electrons starting from terminal 1 (2) will propagate along one zigzag edge and eventually return back to terminal 1 (2) along the same zigzag edge under PMF, see the black arrows in Figs.~\ref{fig:5}(a) and \ref{fig:5}(b), also consistent with the schematic diagram in Fig.~\ref{fig:4}(e). We also give the local occupation number of the two-terminal device [see Fig.~\ref{fig:4}(g)] under the real MF to make a comparison, see Figs.~\ref{fig:5}(c) and \ref{fig:5}(d). Differently, $n_{\bf i}^1$ ($n_{\bf i}^2$) shows that electrons originating from terminal 1 (2) will flow clockwise to terminal 2 (1) along the conductive chiral edge states at the left (right) armchair boundary under the Lorentz force, see Fig.~\ref{fig:5}(c) [Fig.~\ref{fig:5}(d)]. In general, whether the edge state appears at armchair edges can be directly reflected in occupation numbers. The chiral edge states can survive at the armchair edges under the real MF, thus electrons originating from one terminal can go to another by the edge states at armchair edges. However, there is no edge state at the armchair edges under PMF, thus electrons originating from one terminal will finally return to the same terminal by the counterpropagating edge states at zigzag edges.

\section{\label{secS6}A scheme of valley-polarized multi-way switch}

\begin{figure}[ht]
	\includegraphics[scale=0.30]{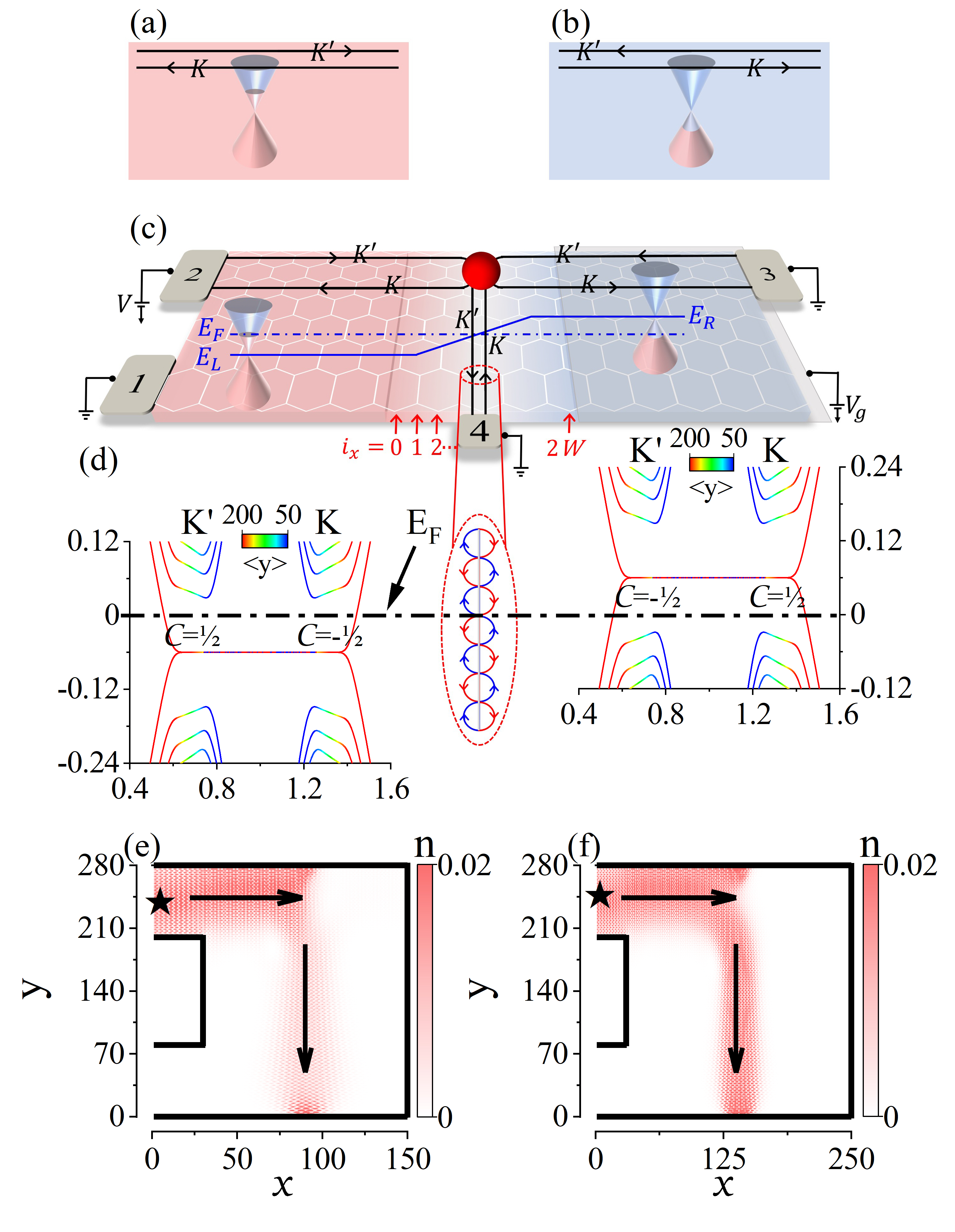}
	\centering
	\caption{A valley-polarized multi-way switch. (a) and (b) show the valley-polarized edge states in an individually strained $n$-type and $p$-type graphene, respectively. Here the edge states only present at the zigzag boundary and there is absence of edge state at the armchair boundary. 
(c) Schematic diagram of a $p$-$n$ junction based on the strained graphene. The $n$-region is depicted in pink and the $p$-region in blue. The corresponding energy bands are shown in middle panel (d) with the on-site energy $E_L=-E_R=0.06t_0$. The red dot in (c) schematically represents the scattering of the edge states at the fork, where the snake states connect the upper and lower zigzag edges, see the enlarged figure for clarity. Electrons are injected from the upper left terminal 2 and can be transmitted to terminal 3 or 4 by adjusting the gate voltage $V_g$ on the right half of the sample. (e) and (f) are the local occupation of electrons vs lattice indices $(x,y)$ for a two-terminal $p$-$n$ junction device with domain wall width $W=0$ and $W=100$, respectively. The width of the $n$-region and the $p$-region is $N_{n}=N_{p}=60$, and the energy is set as $E=E_F=0$. The black star symbols the terminal with a higher voltage. Black solid lines mark the boundary of the device while the black solid arrows indicate the propagation path of electrons under the bias. Electrons can transport from the upper zigzag boundary to the lower one along the interface of the $p$-$n$ junction.}
	\label {fig:6}
\end{figure}

From the above analysis, we know that edge states cannot propagate from the upper zigzag boundary to the lower one in an individually strained graphene with a PMF, due to the absence of edge states at the armchair edges. For quantum Hall effect in graphene, the $p$-$n$ junction, carrying both electrons and holes simultaneously, could allow carriers to propagate at the interface~\cite{Phys.Rev.Lett.101.166806}. Enlightened by this, in this section we investigate the propagation of electrons in a $p$-$n$ junction under strain-induced PMF. Figures~\ref{fig:6}(a) and \ref{fig:6}(b) first show the flow of valley-polarized edge states in two kinds of strained rectangular-shaped graphene nanoflakes, with the $n$-type carriers (electron) for filling factor $\nu_{n}=2$ and $p$-type carriers (hole) for filling factor $\nu_{p}=-2$, respectively. 
From the previous sections, we know that only a pair of valley-polarized counterpropagating edge states survive at the upper zigzag edge in the $n$-region and $p$-region for $\nu=\pm 2$ [see red colors in Fig.~\ref{fig:1}(b)]. However, due to the change of carrier types, electrons and holes feel opposite pseudo-Lorentz force for the same valley, thus the corresponding edge states present opposite chirality at the zigzag boundary. As shown in Figs.~\ref{fig:6}(a) and \ref{fig:6}(b), electrons coming from $K'$ ($K$) valley propagate from the left (right) to the right (left) in the $n$-region while holes coming from $K'$ ($K$) valley propagate from right (left) to the left (right) in the $p$-region.
Here, we emphasize no edge state appears at the armchair edges for both the isolated $n$-type and $p$-type strained graphene. Next, we connect the $n$-region and $p$-region together to construct a $p$-$n$ junction, as shown in Fig.~\ref{fig:6}(c), and a pair of valley-polarized counterpropagating edge states will appear at the interface of the $p$-$n$ junction.
In the central domain wall region, the on-site energy $\varepsilon_{\bf i}$ is varied as $\varepsilon_{\bf i}=i_x(E_R-E_L)/2W+E_L$ with $i_x=0,1,2,\cdots, 2W$. Here $i_x$ denotes the lattice position along the $x$ direction as shown in Fig.~\ref{fig:6}(c) and $W$ represents the domain width. $E_{L}$ and $E_{R}$ are the on-site energy for the $n$-region (pink region) and $p$-region (blue region), respectively. The corresponding energy spectra for the $n$-region and $p$-region with $E_L=-E_R=0.06t_0$ are shown in Fig.~\ref{fig:6}(d). Even though no edge state appears at the armchair edges before, once we connect the two regions together, a pair of valley-polarized counterpropagating edge states intriguingly form at the interface.

In some sense, this graphene $p$-$n$ junction under the PMF is different from the previous one under the real MF. In the latter, edge states always exist at armchair boundaries for an individual graphene nanoflake, thus it will naturally form into interface states once the $p$-$n$ junction is constructed. In the former, edge states are absent at armchair boundaries for an individual graphene nanoflake, but the interface states abruptly appear when $p$-$n$ junction is constructed. In principle, this can be illustrated by valley Chern numbers, $C_{K,K'}$. Due to the opposite PMFs at $K$ and $K'$ valleys, neglecting the spin degeneracy, $C^{n}_{K}=-\frac{1}{2}$ and $C^{n}_{K'}=\frac{1}{2}$ for $n$-region, while $C^{p}_{K}=\frac{1}{2}$ and $C^{p}_{K'}=-\frac{1}{2}$ for $p$-region, see Fig.~\ref{fig:6}(d) with $E_F=0$. Since the valley Chern number $C_{\sigma}$ ($\sigma=K,K'$) for the same valley is opposite at $n$-region and $p$-region, edge states propagate with different directions (chiralities) at the two regions for the same valley.  
At the interface of the $n$-region and $p$-region, the interface states depend on the difference of Chern number between left and right regions (denoted as $C_{\sigma}^d$) across the interface~\cite{Phys.Rev.B.92.041404}.
More specifically, $C_{K}^d=C^{n}_{K}-C^{p}_{K}=-1$ for valley $K$ while $C_{K'}^d=C^{n}_{K'}-C^{p}_{K'}=1$ for valley $K'$. 
Consequently, one interface state from $K'$ valley emerges and propagates from the upper to the lower boundary while another from $K$ valley emerges and propagates from the lower to the upper boundary, also shown in Fig.~\ref{fig:6}(c). In other way, the interface states can also be understood as snake states along the interface~\cite{Phys.Rev.B.105.195420,addref3,Phys.Rev.B.104.125428}, in view that carriers experience the opposite pseudo-Lorentz force in $p$-region and $n$-region and thus walk opposite cyclotron orbits, see the enlarged diagram in Fig.~\ref{fig:6}(d). Such snake states at the interface of the $p$-$n$ junction are also valley-polarized, and can transport electrons from the upper boundary to the lower one.

Our above analysis could be proved by local occupation number of the two-terminal $p$-$n$ junction device with terminals 1 and 2 (first neglecting the terminals 3 and 4), see Fig.~\ref{fig:6}(c). By applying the higher voltage in terminal 2 (marked by the black star) and ground terminal 1, Figs.~\ref{fig:6}(e) and \ref{fig:6}(f) give the local occupation number $n_{\bf i}^{2}$ of the different $p$-$n$ junctions with junction width $W=0$ and $W=100$, respectively. In sharp contrast to Fig.~\ref{fig:5}(b) where electrons always flow along the upper zigzag boundary, electrons originating form terminal 2 first flow along the upper zigzag boundary to the interface of the $p$-$n$ junction, and then mainly turn to the interface snake states rather than keep on moving along the previous path due to the flipping of chirality of edge states across the $p$-$n$ junction. They finally reach the lower zigzag boundary as we expected (see the black arrows). Here the terminal 1 is totally insulated since there is no channel flowing into it. Comparatively, $p$-$n$ junctions with a wider domain wall width $W$ allow electrons flow through interface states more easily [Fig.~\ref{fig:6}(f)], in view of weaker scatterings at the intersection.

The above results have possible interesting applications. In Fig.~\ref{fig:6}(c), we set the left pink region always $n$-region with $\nu_{n}=2$ and the chemical potential in the right blue region can be controlled actually via an additional electrostatic gate. Thus, we can freely manipulate this region to $p$-region or $n$-region. Considering a four-terminal device with terminals 1-4 in Fig.~\ref{fig:6}(c). A voltage is applied at terminal 2 and the rest terminals are grounded. In this case, if we apply a local top gate $V_g$ on the blue region to drive it into $n$-region with $\nu_{n}=2$, the device acts as a $n$-$n$ junction. Because there are only a pair of counterpropagating edge states at the upper zigzag boundary, electrons originating from terminal 2 cannot flow into terminal 4 but must flow into terminal 3 finally, like the result in Fig.~\ref{fig:5}(b). On the other hand, if we apply a local top gate $V_g$ on the blue region to drive it into $p$-region with $\nu_{p}=-2$, the device acts as a $n$-$p$ junction as shown in Fig.~\ref{fig:6}(c). In this case, electrons originating from terminal 2 hardly flow into terminal 3 but mainly flow into terminal 4, like the result in Fig.~\ref{fig:6}(e). Therefore, using a local gate on the strained graphene, we can control the current from the terminal 2 to terminal 3 or 4, acting like a single pole double throw switch. Notably, these channels are all valley-polarized, which are perfectly transparent for one valley but completely opaque for another along a specified direction. This makes it possible for the above $p$-$n$ junction to conduct valley-polarized currents, behaving as a perfect valley filter~\cite{Nat.Phys.3.172,Phys.Rev.Lett.106.136806}, which may be valuable for future valleytronics applications~\cite{Science.362.1149}.

\section{\label{secS7}Conclusions}
In summary, we investigate the distribution of conducting edge states at distinct edges in graphene under a PMF induced by the inhomogeneous uniaxial strain. Quite different from the usual case under a real MF, even though pseudo-Landau levels emerge in the bulk, the counterpropagating edge states are found to propagate only along zigzag edges in strained graphene but are absent at the armchair edges due to the valley mixing. Therefore, the armchair edges are insulated in strain-induced quantum valley Hall effect, which can be well shown by the wave function distributions, zero transmission coefficients, and local occupations. Considering electrons cannot transport from one zigzag edge to another in an individually strained graphene, we propose that the interface snake states could appear in a $p$-$n$ junction with PMF induced by strain, which originate from the difference of the valley Chern numbers in $p$-region and $n$-region. 
This $p$-$n$ junction could achieve a single pole double throw switch just via a local gate control, and also act as a valley filter.

\section*{ACKNOWLEDGMENTS}
This work was financially supported by the National Natural
Science Foundation of China (Grant No. 12374034 and No. 11921005), 
the Innovation Program for Quantum Science and Technology
(2021ZD0302403), and the Strategic priority Research
Program of Chinese Academy of Sciences (Grant No.
XDB28000000). We also acknowledge the Highperformance
Computing Platform of Peking University for
providing computational resources.

\section*{APPENDIX A: Wave function centers of the graphene nanoflakes}

\begin{figure*}
	\includegraphics[scale=0.5]{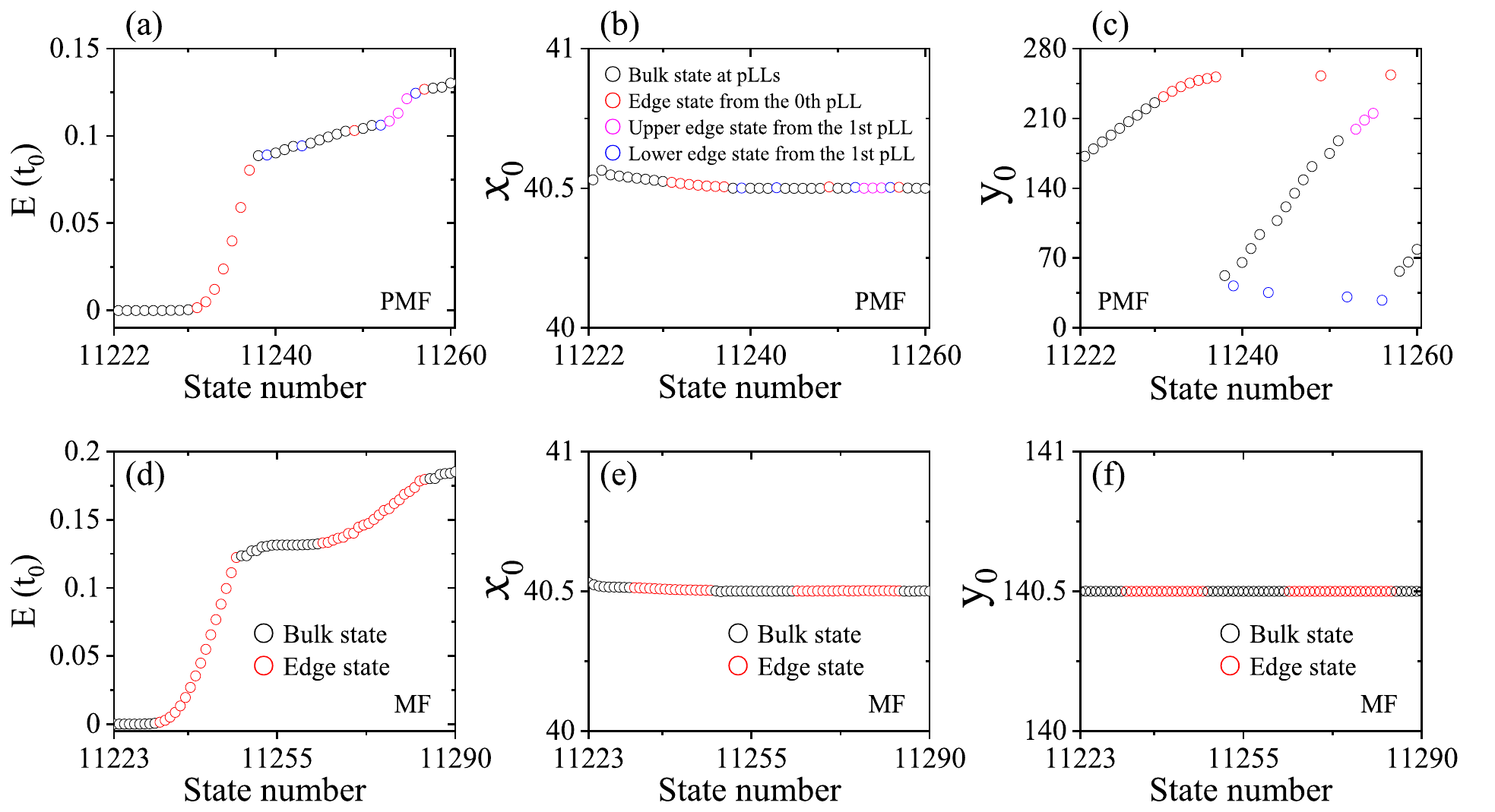}
	\centering
	\caption{(a) [(d)] is the enlarged figure of the region surrounded by the green dashed lines in Fig.~3(a) [Fig.~3(g)]. (b),(c) [(e),(f)] show the wave function centers $x_0$, $y_0$ corresponding to the states in (a) [(d)], respectively. The word pLL in the legend in (b) stands for pseudo-Landau level. Here the legend in (b) is shared for (a)-(c).}
	\label {fig:A1}
\end{figure*}

To understand the calculated low-energy states in Figs.~\ref{fig:3}(a) and \ref{fig:3}(g) comprehensively, we also show their wave function centers by $x_0=\int x|\Psi(x,y)|^2 dr$ and $y_0 = \int y|\Psi(x,y)|^2 dr$. Since the results of $E\leq 0$ is symmetric to those of $E\geq 0$, we only concentrate on the low-energy states with $E\geq 0$. And the regions surrounded by the green dashed lines in Figs.~\ref{fig:3}(a) and \ref{fig:3}(g) are enlarged and shown in Figs.~\ref{fig:A1}(a) and \ref{fig:A1}(d). For the case of PMF, we can clearly mark the bulk states from pseudo-Landau levels all by black circles and edge states from different pseudo-Landau levels by red, magenta and blue circles, respectively [see legends in Fig.~\ref{fig:A1}(b)]. They can be well-distinguished by plotting their density profiles (not shown here) and also by their values of $y_0$ in Fig.~\ref{fig:A1}(c). Note that within the range about $E\in[0.09t_0,0.11t_0]$, we can find the coexistence between edge states from the zeroth pseudo-Landau level (red circles) and lower edge states from the first pseudo-Landau level (blue circles) as well as bulk states from the first pseudo-Landau level (black circles), which roughly corresponds to the range of the tilted first pseudo-Landau levels between $\upsilon = 2$ and $\upsilon = 6$ in Fig.~\ref{fig:1}(b).  In Fig.~\ref{fig:A1}(b), we can see that $x_0\approx 40.5$ for all the states reflecting the mirror symmetry in the $x$ direction with $N_x=80$. While in the $y$ direction in Fig.~\ref{fig:A1}(c), the introduced strain breaks the mirror symmetry, thus $y_0$ is no longer located at the center along the $y$ direction. In detail, with the increase of the state number, we can see a periodic tendency where $y_0$ moves across the bulk (black circles) to the upper edge (red and magenta circles). This reflects the transition from the pseudo-Landau level to upper edge states as the energy climbs [like transitions from green color to red color in Fig.~\ref{fig:1}(b)]. Besides, $y_0$ moving towards the lower edge (blue circles) also indicates the appearance of lower edge states for higher pseudo-Landau levels [blue color lines in Fig.~\ref{fig:1}(b)]. 

For a comparison, we also show the corresponding results under the real MF in Figs.~\ref{fig:A1}(d)-(f). Since the Landau levels are all flat and all edge states will form a closed loop in the graphene nanoflakes, we only distinguish the bulk states (black circles) and edge states (red circles), see Fig.~\ref{fig:A1}(d). We can see that $x_0$ and $y_0$ are always distributed at the center of the system with $x_0\approx 40.5$ and $y_0\approx 140.5$ [Figs.~\ref{fig:A1}(e) and \ref{fig:A1}(f)], since the mirror symmetries both along the $x$ and $y$ directions are kept here with $N_x=80$ and $N_y=280$. The different characteristics of $y_0$ for PMF and MF are well-consistent with previous results and indicate the distinctions in distributions for two types of edge states.

\end{document}